# Synthesis, crystal growth and characterization of the pyrochlore $Er_2Ti_2O_7$

Q. Wang[a,b], A. Ghasemi[a], A. Scheie[a], S.M. Koohpayeh[a,*]

[a]*Institute for Quantum Matter, Department of Physics and Astronomy, Johns Hopkins University, Baltimore, MD 21218, USA*

[b]*Institute of Laser Engineering, Beijing University of Technology, Beijing 100124, China*

**Abstract:** Pyrochlore erbium titanate samples in the form of powders (with nominal compositions of $Er_2Ti_{2+x}O_7$, $-0.08 \leq x \leq 0.08$) and bulk single crystals were prepared to elucidate the effect of synthesis and growth conditions on their composition, structure, and transition temperature. All samples were characterized using X-ray diffraction and specific heat measurements. Larger lattice parameters were measured for the Ti-deficient stuffed powders, $Er_2(Ti_{2-x}Er_x)O_{7-\delta}$, while Ti-rich anti-stuffed powders, $(Er_{2-x}Ti_x)Ti_2O_{7+\delta}$, showed a decrease in lattice parameter. The lattice parameter for the stoichiometrically synthesized powder, $Er_2Ti_2O_7$, was measured to be $a$ = 10.07522(9) Å. Single crystals grown by the conventional floating zone (FZ) technique were Ti deficient (stuffed) and darker in color due to oxygen vacancies. Using the traveling solvent floating zone (TSFZ), a high structural quality, transparent and stoichiometric $Er_2Ti_2O_7$ single crystal was grown at a lower temperature using the $TiO_2$ solvent. Heat capacity measurements of the TSFZ grown $Er_2Ti_2O_7$ crystal and stoichiometric powder exhibited a very close magnetic transition temperature $T_C$ ~1.23K that was suppressed in off-stoichiometric samples, demonstrating correlations between stoichiometry and ground state magnetism in $Er_2Ti_2O_7$.

---

[*]Corresponding author: S.M. Koohpayeh (Koohpayeh@jhu.edu)

**1. Introduction**

Geometrical frustrated pyrochlore structures ($A_2B_2O_7$) have attracted considerable attention over the past decades due to their exotic magnetic properties, *e.g.* spin ices,[1] spin liquids,[2] and noncolinear XY orders.[3-6] The pyrochlore lattice is a cubic structure with a basis of corner-sharing tetrahedra where the A and B ions reside at the corners of three-dimensional tetrahedral networks. If either the A or B site is occupied by a magnetic atom with an antiferromagnetic nearest-neighbor interaction, then the tetrahedral structure can be subject to strong geometric magnetic frustration.[7,8] Such is the case for rare-earth titanate $Er_2Ti_2O_7$, which has been identified as an *XY* antiferromagnet.[9]

$Er_2Ti_2O_7$ is an antiferromagnetic insulator with a negative Curie-Weiss temperature ($\theta_{CW}$ ~22 K).[10,11] Ordering occurs near $T_N \sim 1.2$ K, $Er_2Ti_2O_7$ undergoes a continuous phase transition to a non-coplanar $\psi_2$ antiferromagnetic ordered state,[3,12-14] where the $Er^{3+}$ spins with *XY* anisotropy are energetically constrained to lie in the plane perpendicular to the local <111> axes due to the dominating $J_{\{+/-\}}$ coupling $S_i^{\{+/-\}}$ and $S_j^{\{-/+\}}$.[15] However, it is still controversial whether the selection mechanism for the $\psi_2$ state is order-by-disorder[15-18] or an energetic selection by virtual crystal field transitions.[19-21] The low-temperature magnetism is expected to be particularly sensitive to small perturbations, including those of defects, possibly resulting in discrepancies between experimental results.

There have been numerous studies of $Er_2Ti_2O_7$ powders and single crystals in the past decades. However, inconsistencies exist in the experimental results, most notably in reported lattice parameters which range from 10.0450 Å to 10.0869 Å depending on the preparation method used (*e.g.* solid-state reaction, floating zone technique, flux technique, *etc.*).[22-29] In other pyrochlore materials, small changes in composition have been shown to dramatically impact the ground state magnetism, as reported for $Tb_{2+x}Ti_{2-x}O_{7+y}$,[30] $Yb_2Ti_2O_7$,[31] $Pr_{2+x}Zr_{2-x}O_{7-x/2}$ (0.98≤Pr/Zr≤1.02),[32] and $Eu_2Ir_2O_7$.[33,34]

Here, through a systematic study, we investigate the effects that synthesis and growth conditions can have on the stoichiometry and structural and physical properties. We prepared single phase powders with nominal compositions of $Er_2Ti_{2+x}O_7$ (-0.08≤$x$≤0.08) and examined the formation of site disordering, homogeneity and stoichiometry. In this work, we also demonstrate the process to grow a high-quality and stoichiometric $Er_2Ti_2O_7$ single crystal by the traveling solvent floating zone technique (TSFZ), which is reported for the first time. We finally use low-temperature specific heat to show how off-stoichiometry and defects (e.g. stuffing and anti-stuffing) affect the bulk properties.

## 2. Experimental procedures

### 2.1. Synthesis of powder samples

Polycrystalline $Er_2Ti_{2+x}O_7$ (-0.08≤$x$≤0.08) samples were synthesized via a solid-state reaction using high purity powders of $Er_2O_3$ (99.99% Alfa Aesar) and $TiO_2$ (99.99% Alfa Aesar) in nominal ratios. The starting materials were dried overnight at 1000 °C, and then weighed to the appropriate ratio. The samples were heated several times at 1400 °C, and additionally at 1500 °C, in air with an intermediate grinding.

### 2.2. Crystal growth of the $Er_2Ti_2O_7$

Single crystal growths were performed using the stoichiometrically synthesized feed rods (of typically 5 mm in diameter and 80 mm in length). A four-mirror optical image furnace (Crystal Systems, Inc., FZ-T-4000-H-VII-VPO-PC) equipped with four 1 kW halogen lamps as the heating sources was used to grow the crystals using the float-zone techniques.[35] The seed and feed rods were connected to the lower and upper shafts respectively, while the melt-zoning was performed by moving the heating source (lamps) upwards.

### 2.3. Characterization

The powder X-ray diffraction patterns were obtained at room temperature over a 2θ range from 5° to 120° using a Bruker D8 focus diffractometer operating with Cu Kα radiation and a LynxEye detector. Rietveld refinements were performed using TOPAS software (Bruker AXS) to determine the lattice parameters. Back-reflection X-ray Laue diffraction was used to check the orientation and crystalline quality of the crystals. The Raman scattering signals were captured by a high-resolution confocal Raman spectrometer (Horiba/Jobin-Yvon T64000) using the 514.5 nm line of the Spectra-Physics $Ar^+$-$Kr^+$ laser for excitation.

The heat capacity ($C_p$) measurements were carried out in zero magnetic field using a Quantum Design physical property measurement system (PPMS) over the temperature range 0.4–2.8 K. We measured six samples: two single crystals and four powders. The two crystals were cut and polished into the small thin plates of masses 1.59 mg (TSFZ crystal) and 1.07 mg (FZ crystal). Powder samples of x = -0.16, -0.08, 0.0, and +0.08 were pressed into small pellets of 0.6–0.8 mg and then sintered at 1000 ºC to make them compact. Apiezon N grease was applied to provide thermal contact between the sample and the platform during the measurement. Heat capacity data were collected using semi-adiabatic method and long-pulse method from the same measurement (with the exception of the x = -0.16 powder, where only semi-adiabatic data was collected). The LongHCPulse software was used to compute heat capacity with the long-pulse method which is more sensitive to first-order transitions and self-consistent.[36] The differences in the data obtained by these two methods have been described in our previous work.[31]

## 3. Results and discussion
*3.1. polycrystalline samples*

The polycrystalline samples of $Er_2Ti_{2+x}O_7$ (x = -0.08, -0.04, 0, 0.04, 0.08) were initially sintered several times at 1400 ºC in air. This procedure yielded near-stoichiometric powders, but left noticeable second phases in the material. Using powder X-ray diffraction, $Er_2TiO_5$ second phase (of a fluorite-type structure) was detected for the Ti deficient (x<0) powders, as also predicted in the temperature-composition phase diagram of $Er_2O_3$-$TiO_2$.[37] No indication of second phases was observed in the XRD pattern of the stoichiometrically synthesized powder (x = 0); however, Raman spectrum taken from this powder revealed signs of the $TiO_2$ second phase, as shown in Fig. 1.[38-40] Since other Er rich phases (e.g. $Er_2O_3$) were not seen in both the XRD pattern and Raman spectrum of this powder, the extra small amount of the unreacted $TiO_2$ should have left some levels of Ti deficiencies (or slight stuffing) in this powder. Lattice parameters measured for all the powders synthesized at 1400 °C were very close, ranging from 10.07750(9) Å and 10.07681(7) Å (for x = -0.08 and -0.04 respectively), to 10.07620(9) Å (for x≥0), indicating that stuffing and anti-stuffing levels are not significant. These powders, however, are not pure single phase due to the presence of the second phases of $Er_2TiO_5$ (for x<0) and $TiO_2$ (for x≥0).

To eliminate the second phases, the powder samples of $Er_2Ti_{2+x}O_7$ (-0.08≤x≤0.08) were additionally sintered at a high temperature of 1500 °C in air, and this led to single phase samples with no evidence of secondary phase inclusions. This is consistent with the reported phase diagram[37] which shows a higher solid solubility at temperatures near 1500 °C and above. Lattice constants for all the powders were measured via Rietveld refinements, as reported in Table 1. The lattice parameter $a$ increased for the Ti deficient powders (x<0) due to stuffing or replacement of $Ti^{4+}$ cations with the larger $Er^{3+}$ cations, as presented in the form of $Er_2(Ti_{2-x}Er_x)O_{7-\delta}$. For the Ti rich powders (x>0), however, the lattice parameter $a$ decreased because of anti-stuffing or replacement of $Er^{3+}$ cations with the smaller $Ti^{4+}$ cations, $(Er_{2-x}Ti_x)Ti_2O_{7+\delta}$. Although we could not precisely resolve the levels of stuffing or anti-stuffing with X-ray refinements, we observe that the lattice parameters vary systematically with composition, and there is no sign of second phases in the XRD patterns. Therefore, we assume that the stuffing and anti-stuffing have fully occurred, as presented based on the Ti concentration, x, in Table 1.

A post-reaction annealing in oxygen was found to be necessary to accomplish the anti-stuffing process for the Ti-rich compounds (x>0), as similarly reported for the Ti-rich holmium titanate compounds.[41] This is shown in Fig. 2a where an extra small shoulder (peak) of a similar type pyrochlore phase was detected in the XRD pattern for the x = +0.08 powder, which was then corrected by annealing in oxygen at 1300 °C. This is consistent with the charge neutrality of the anti-stuffed compounds, $(Er_{2-x}Ti_x)Ti_2O_{7+\delta}$, for which an oxygen rich atmosphere is needed for the structure to be developed. However, for the stuffed compounds, $Er_2(Ti_{2-x}Er_x)O_{7-\delta}$, oxygen deficient environments (such as air and argon) are more favorable, while synthesizing in oxygen rich atmospheres can disrupt the structure. This was examined by annealing the stuffed powder (of x = -0.04) in oxygen which led to get an extra peak (shoulder) in the XRD pattern, as shown in Fig. 2a. The measured lattice constants, reported in Table 1 for both the anti-stuffed (x>0) and stuffed (x<0) compounds, did not change noticeably after annealing in oxygen, and this is supported by identical peak positions shown in Fig. 2a. The color of these powders also appeared to be the same before and after annealing in oxygen.

Synthesizing at a higher temperature of 1500 °C in air for 30-40 h (with intermediate grindings) also helped to complete the reaction for the stoichiometric powder (x = 0) and obtain a pure single phase with the measured lattice constant of $a$ = 10.07522(9) Å. Furthermore, it was found that over sintering at the high temperature of 1500 °C led to a small gradual reduction of the lattice constant. As shown in Fig. 2b, additional five times heating cycles decreased the lattice constant by 0.03%, from $a_1$ =10.07522(9) Å to $a_6$ =10.07231(8) Å, and the attained powders seemed to be slightly darker in color. A series of post-annealing cycles in oxygen (with intermediate grindings) steadily increased the lattice constant to its original value. Extra small peaks (of similar intensity as was seen in Figs. 2a) also appeared, and then disappeared in the XRD during the process of oxygen annealing. This seems to indicate that over sintering at high temperatures in air introduces anion (oxygen) vacancies which were then corrected by oxygen annealing, and this process was found to be accompanied with some structural modification, as thoroughly reported in reference.[42]

The lattice parameter $a$ = 10.07522(9) Å is proposed for the stoichiometric pyrochlore $Er_2Ti_2O_7$. Fig. 3(a) shows the excellent fit between the standard $Fd\bar{3}m$ model for pyrochlores and the high-quality diffraction pattern of the stoichiometric $Er_2Ti_2O_7$. A color comparison for the powder products is also provided in Fig. 3b, showing a change from dark-pink (for x = -0.08) to light-pink (x = 0), and then to deep-pink (for x = +0.08).

*3.2. Crystal growth*

Our first attempts to grow erbium titanate single crystals by the conventional optical floating zone (FZ) method using the stoichiometric polycrystalline rods yielded a flurry of problems. The as-grown crystal was dark-pink in color due to the change of Ti oxidation state and formation of $Ti^{3+}$ cations, as similarly seen for $Yb_2Ti_2O_7$,[31] $Ho_2Ti_2O_7$,[41] and $Pr^{4+}$ inclusion in $Pr_2Zr_2O_7$ single crystals.[32] A high concentration of $TiO_2$ impurity phase (~12.5% in mass) was also identified in the XRD taken from the molten zone. This indicates that the grown crystal is Ti deficient. This was confirmed by larger lattice parameters measured along the crystal. The lattice parameter *a* varied considerably from 10.08901(9) Å, measured from a crystal piece at the initial stage of growth (beginning of the grown crystal), to 10.08130(5) Å, at the end of the crystal, as shown in Fig. 4. Based on the trend of lattice constant shown in Fig. 4, these larger values correspond to Ti deficiencies levels of about 5.5% ($x = -0.11$) at the beginning and 2.6% ($x = -0.52$) at the end of the float-zone crystal. No vaporization of other phases was detected during the FZ growth process.

All these features suggest that $Er_2Ti_2O_7$ does not melted congruently and the grown crystal is not stoichiometric. Therefore, the traveling solvent floating zone (TSFZ) technique[43] was applied to grow this compound at a lower growth temperature using a $TiO_2$ rich solvent (30wt% $TiO_2$ and 70wt% $Er_2Ti_2O_7$), as similarly used for the growth of $Yb_2Ti_2O_7$ and $Ho_2Ti_2O_7$ single crystals.[31,41] During the growth, the molten zone was passed upwards at a rate of 0.5 mm/h. Rotation rates of 3 and 6 rpm were employed in opposite directions for the feed rod and the growing crystal, respectively. Crystal growth was carried out at a power level of 63.8%, under a dynamic oxygen atmosphere with a pressure of 1 atm and a flow rate of 10 mL/min. A high quality, transparent and pink single crystal, shown in Fig. 5, was obtained. X-ray Laue pictures taken at regular intervals along the length of the crystal indicated that it exhibited an equally high crystalline quality with no detectable variation of the orientation between pictures and no evidence of spot splitting or distortion, as pictured in Fig. 5b. (The X-ray beam used had a diameter of 1 mm and allowed orientation variations of less than 1° to be detected). Unlike the FZ grown crystal, the quality and purity of the TSFZ crystal is also confirmed by the consistent and unchanged lattice parameter of 10.07525(6) Å which is identical to the stoichiometrically synthesized powder, $a = 10.07522(9)$ Å, reported in Table 1. No signs of stuffing or anti-stuffing and color changes were seen in the TSFZ crystal, and this additionally confirms that the grown crystal is stoichiometric having a high crystalline quality.

*3.3. Heat capacity*

Fig. 6 shows low-temperature ($T$ = 0.4–2.8 K) specific heat capacity $C_P$ measurements of single crystals and powder samples ($x$= -0.16, -0.08, 0.0, 0.08). All $C_P$ ($T$) data exhibit a sharp peak, corresponding to a second-order phase transition into a magnetically ordered state, falling within the temperature range of ~1.18–1.24 K. The TSFZ crystal exhibits a high $T_c$ at 1.23(2) K (Fig. 6a) in close agreement with the stoichiometric powder of 1.23(7) K. Meanwhile, the off-stoichiometric samples show a suppressed transition temperature as shown in Fig. 6(b), with the TSFZ crystal being the sharpest. Fig. 6(c) shows the transition temperature plotted against stuffing level, demonstrating that slight stuffing and anti-stuffing levels suppress the ordering temperature by several percent. It has been previously observed that large levels of stuffing ($x$>0.17) suppress the ordering temperature,[12] but we observe the effect even for small levels. The high $T_c$ at 1.23(2) K shown in this study is also higher than the reported transition temperatures, 1.173(2) K and 1.1 K, measured from the FZ grown crystals.[3,44]

The dependence of transition temperature upon defect levels resembles the dramatic dependence observed in $Yb_2Ti_2O_7$.[31,45-47] Although the observed dependence in $Er_2Ti_2O_7$ is weaker than $Yb_2Ti_2O_7$, it is still true that sharpest heat capacity peak corresponds to the cleanest sample, and thus, the ground state of $Er_2Ti_2O_7$ does depend upon quenched disorder. A qualitative explanation for the difference in sensitivity to disorder between these two compounds is that $Er_2Ti_2O_7$ lie far away from competing XY phases while $Yb_2Ti_2O_7$ is proximity to competing XY phases.[48]

**4. Conclusions**

Formation of site disordering, vacancy defects, phase purity and structural quality in a series of erbium titanate powders were found to be significantly affected by synthesis conditions and small deviations from stoichiometry. In Ti-deficient stuffed powders lattice parameter *a* increases relative to the stoichiometric powder, while it decreases for the anti-stuffed Ti-rich powders. Erbium titanate was found to melt incongruently, and as a result, the float-zone (FZ) grown crystals were stuffed and non-stoichiometric due to both Ti and O vacancies. Growth of a stoichiometric and high crystalline quality of $Er_2Ti_2O_7$ single crystals was successfully performed using the traveling solvent floating zone (TSFZ) technique. The high heat capacity $T_C$ at 1.23(2) K of the TSFZ grown crystal was in close agreement with the stoichiometric powder of 1.23(7) K, while the $T_C$ were suppressed by several percent in the off-stoichiometric powders or FZ crystal samples, indicating a noticeable dependence of the magnetic ground state upon quenched disorder.

**Conflicts of interest**

There are no conflicts of interest to declare.

**Acknowledgments**

This work was supported as part of the Institute for Quantum Matter, an Energy Frontier Research Center funded by the U.S. Department of Energy, Office of Science, Office of Basic Energy Sciences under Award Number DE-SC0019331. A. Scheie was supported through the Gordon and Betty Moore foundation under the EPIQS program GBMF4532. This work was partly supported by the China Scholarship Council (No.201706540009) during Q. Wang's visit at the Johns Hopkins University in USA. The author would also like to acknowledge C. Wan, H. Man, and N. Drichko for helpful discussions.

TABLE 1. The lattice parameters $a$ of $Er_2Ti_{2+x}O_7$ (-0.08≤$x$≤0.08) powder samples synthesized at 1500℃ in air.

| $x$ | $a$ (Å) | Compounds |
|---|---|---|
| -0.08 | 10.08442(1) | $Er_2(Ti_{2-x}Er_x)O_{7-\delta}$ |
| -0.04 | 10.07998(4) | $Er_2(Ti_{2-x}Er_x)O_{7-\delta}$ |
| 0 | 10.07522(9) | $Er_2Ti_2O_7$ |
| 0.04 | 10.07475(6)* | $(Er_{2-x}Ti_x)Ti_2O_{7+\delta}$ |
| 0.08 | 10.07395(4)* | $(Er_{2-x}Ti_x)Ti_2O_{7+\delta}$ |

* Additionally annealed at 1300℃ in $O_2$.

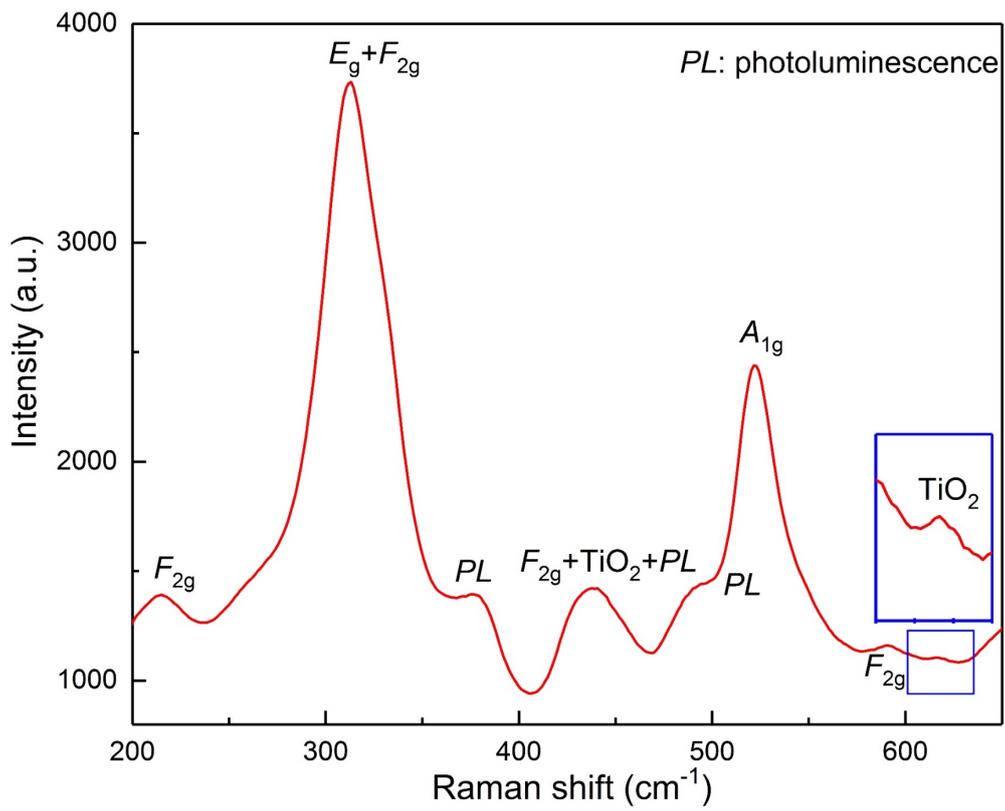

Fig.1. Raman spectra from the stoichiometrically synthesized powder (x = 0) at 1400℃ in air, showing a slight level of rutile $TiO_2$ impurity.

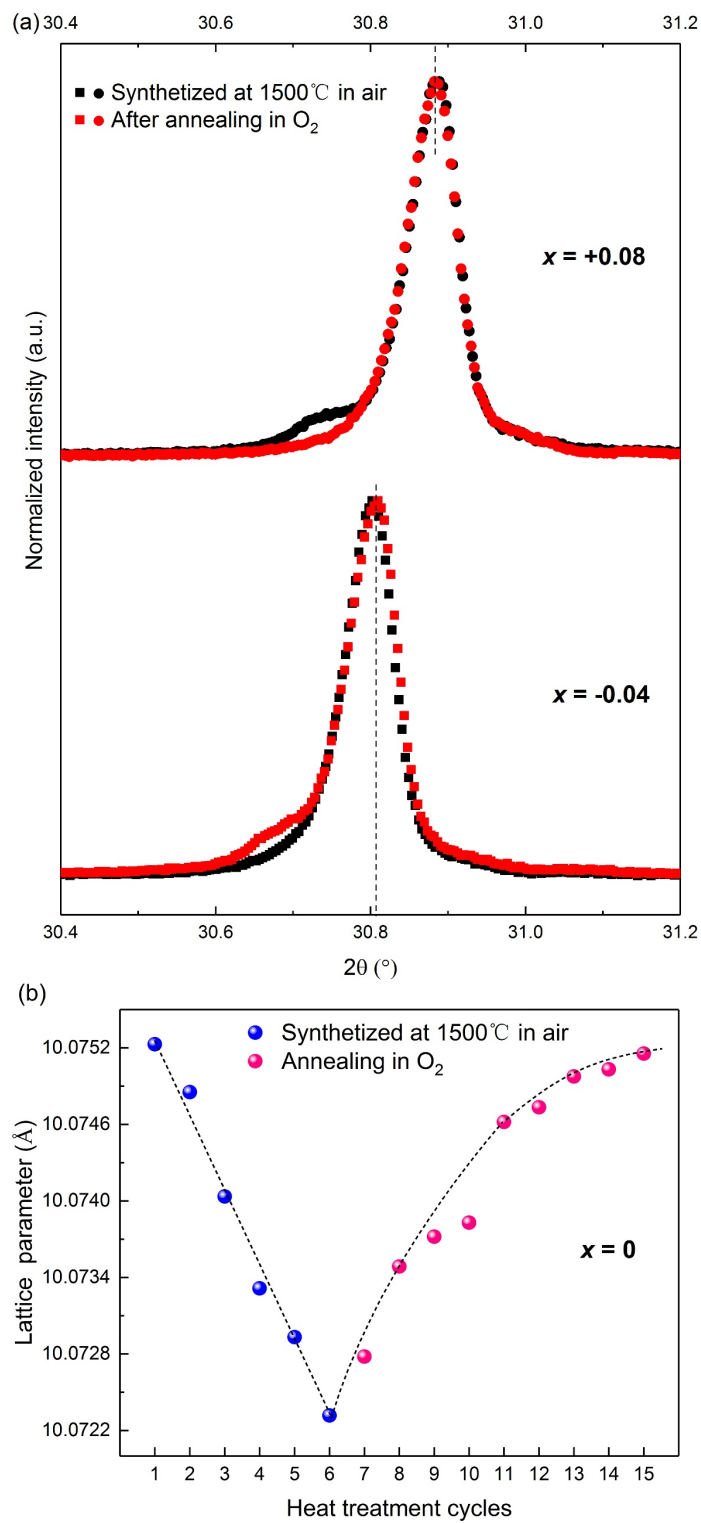

Fig. 2. (a) The shoulder (extra peak) observed in the XRD patterns taken from the powders of $x = 0.08$ (as synthesized at 1500℃ in air) and $x = -0.04$ (after annealing at 1300℃ in $O_2$). (b) Changes of the lattice parameter for the stoichiometrically synthesized powder due to over sintering in air, and post-annealing in oxygen. The dashed line is a guide to the eye.

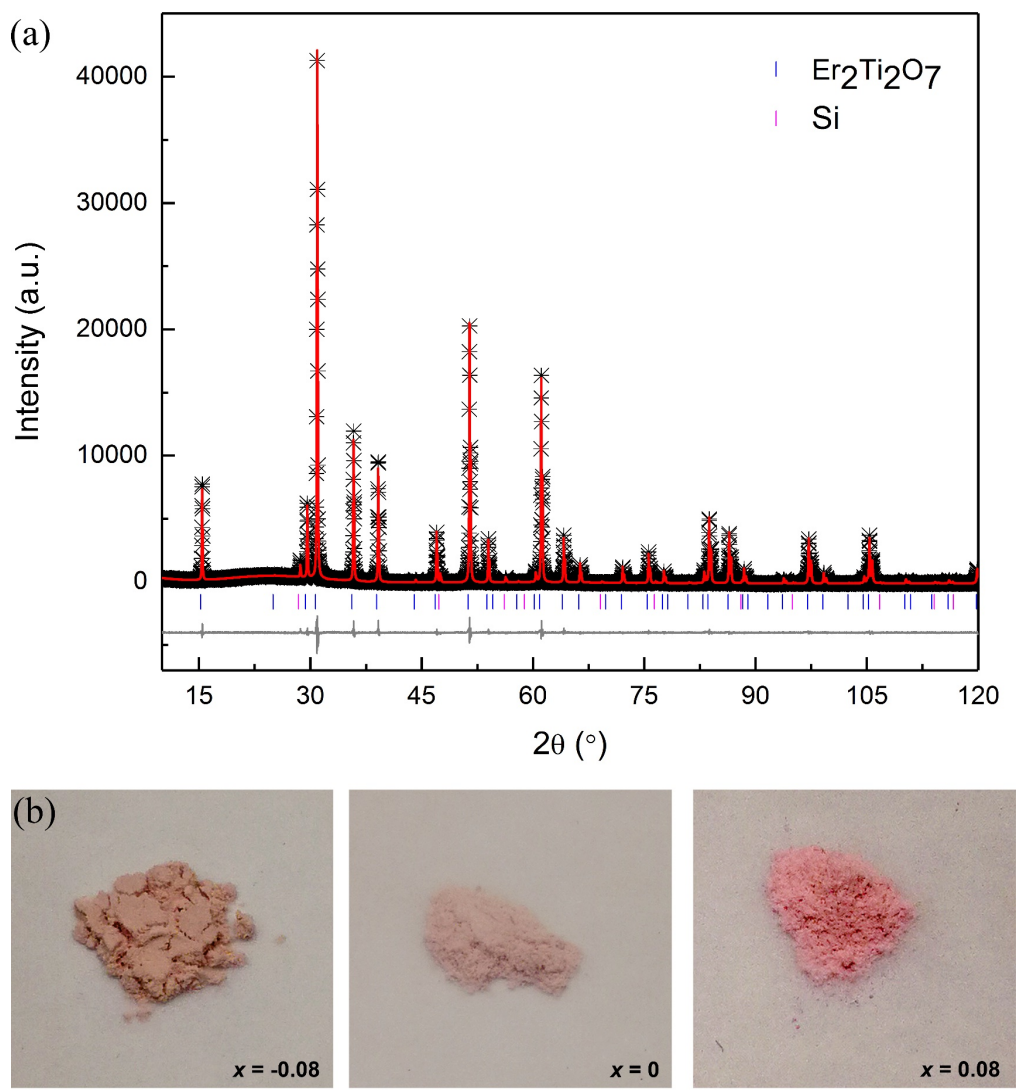

Fig. 3. (a) Rietveld structure refinement of the stoichiometric $Er_2Ti_2O_7$ powder (with raw data in black, fitted pattern in red, difference in gray, and *hkl* reflections in pink and blue). (b) Picture of the $Er_2Ti_{2+x}O_7$ powders at different doping levels, x = -0.08, 0.0, and 0.08, exhibiting a continuous change from dark-pink to light-pink, and then to deep-pink with the increase of x.

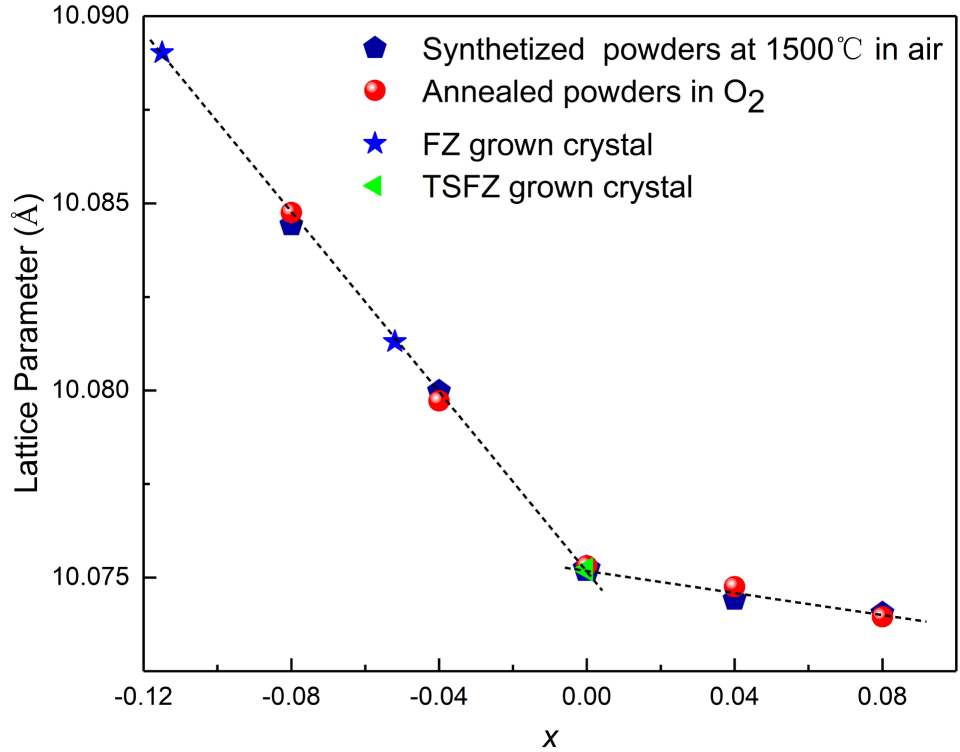

Fig. 4. The lattice parameter trend of polycrystalline $Er_2Ti_{2+x}O_7$ (x = -0.08, -0.04, 0, 0.04, 0.08) powders and single crystals. About 5.5at% (x = -0.11) and 2.6at% (x = -0.052) Ti-deficiencies were estimated at the beginning and end of the FZ grown crystal, respectively. The lattice parameter of the TSFZ grown crystal is identical to the stoichiometrically synthesized powder.

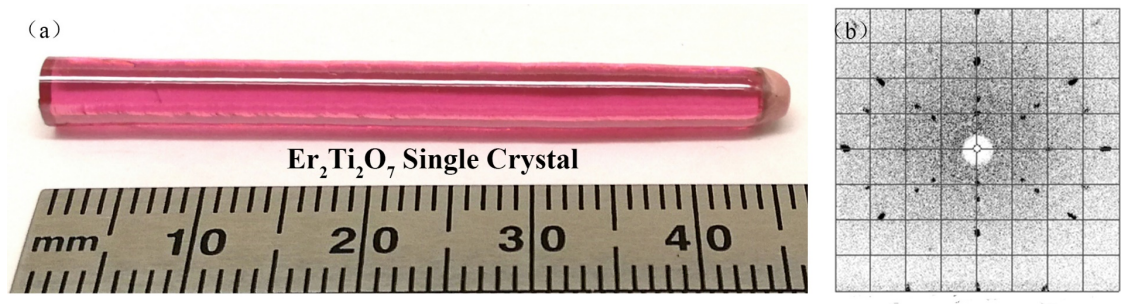

Fig. 5. (a) Image of the Er$_2$Ti$_2$O$_7$ crystal grown by the TSFZ technique, and (b) corresponding Laue diffraction pattern.

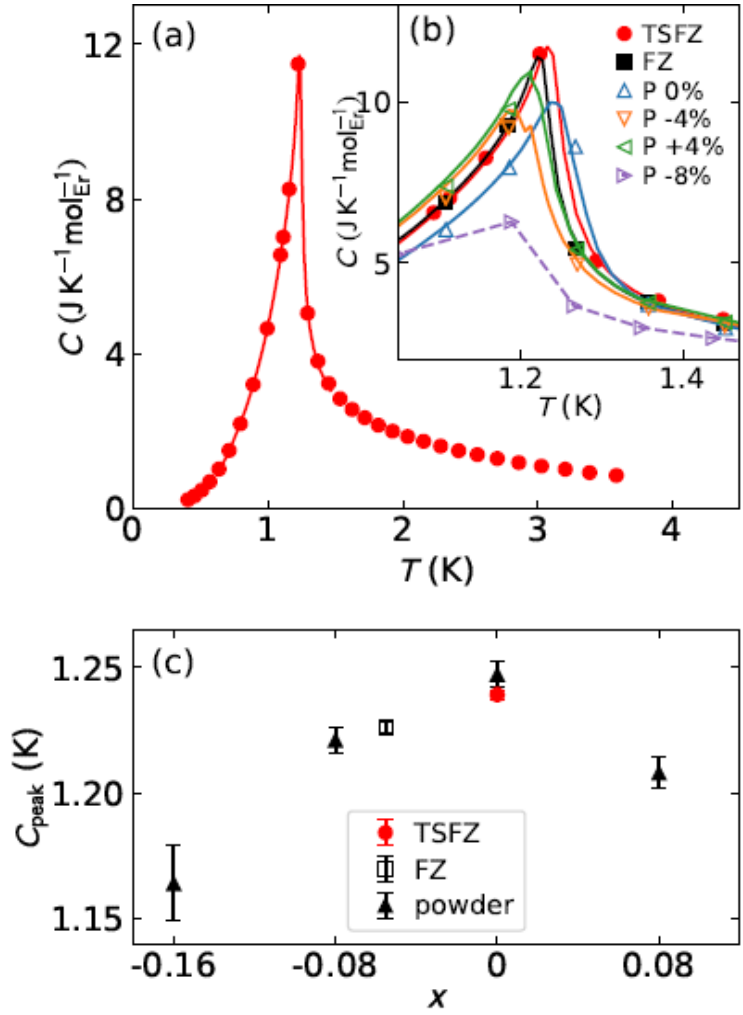

Fig. 6. (a) Low-temperature specific heat of the TSFZ grown $Er_2Ti_2O_7$ crystal, showing a transition at 1.23 K. (b) Heat capacity peaks of $Er_2Ti_{2+x}O_7$ powders and single crystals in detail. The TSFZ grown crystal and stoichiometric powder exhibited a very close transition temperature $T_C$ ~1.23 K that was suppressed in off-stoichiometric samples. Data taken by the semi-adiabatic method are plotted with symbols, and data taken with the long pulse method are plotted with solid lines. (c) Transition temperature as a function of stuffing for the six measured samples.